\documentclass[reprint,aps,twoside,pre,showkeys,showpacs,superscriptaddress]{revtex4-2}
\usepackage{epsfig}
\usepackage{amsmath}
\usepackage{amssymb}
\usepackage{txfonts}
\usepackage{bm}
\def\<{\left\langle}
\def\>{\right\rangle}
\begin{document}
\setcounter{page}{830}
\pagestyle{myheadings}
\markboth{\hfill\rm J\"URGEN STRUCKMEIER\hfill\underline{54}}%
{\underline{54}\hfill\rm CONCEPT OF ENTROPY IN THE REALM OF CHARGED %
PARTICLE BEAMS\hfill}
\title{Concept of entropy in the
realm of charged particle beams}
\author{J\"urgen Struckmeier}
\affiliation{Gesellschaft f\"ur Schwerionenforschung (GSI),
Planckstr.~1, 64291~Darmstadt, Germany}
\date{Received 30 November 1995}

\begin{abstract}
Stochastic phenomena occurring within charged particle beams
can be handled using the Vlasov-Fokker-Planck generalization
of the Vlasov equation.
In particular, this non-deterministic approach can deal with
effects due to Coulomb scattering between the beam particles.
Moreover, stochastic phenomena also occur
in computer simulations of charged particle beams.
Both processes --- although different in their physical nature ---
can be described by the Vlasov-Fokker-Planck equation, since
in both cases the underlying stochastic process can be
classified as a Markov process.
This description is applied to beams in periodic focusing systems.
We derive an equation relating the change of the
$\mu$-phase space entropy to the change of rms-emittance and
``temperature weighted excess field energy''.
This equation enables us both to improve our capability to
interpret the results of computer simulations, as well as to
identify the conditions needed to minimize scattering induced
degradation of the quality of beams circulating in storage rings.
\end{abstract}
\maketitle
\section{INTRODUCTION}
Analytical approaches to particle motion
that are based on the Vlasov equation require
that Liouville's theorem --- though strictly valid only in the
$6N$-dimensional $\Gamma$-phase space --- also applies to the
$6$-dimensional $\mu$-phase space, at least to a good approximation.
This is obviously correct in a regime where the motion of single
particles can be treated as being independent of each other.
In the more realistic case of interacting particles, the Liouville
theorem remains fulfilled in the $\mu$-phase space as long as the
space charge fields can be regarded as smooth macroscopic functions.
This is no longer true for cases where forces
between individual particles play a role.
One example for this ``non-Liouvillean'' behavior is the scattering
induced emittance growth effect within ion beams (``intra-beam scattering'').
A second example is the appearance of numerical noise phenomena
in computer simulations of charged particle beams.
As will be demonstrated in this article, the main source for this
computer artifact originates in the modeling of a real beam,
for performance reasons, by a ``simulation'' beam
containing several orders of magnitude fewer particles.

In order to analyze non-Liouvillean phenomena, a generalization
of the Vlasov equation becomes necessary.
In this article we follow the approach applied earlier by
Chandrasekhar\cite{chandra}, who modeled the non-Liouvillean
contributions to the dynamics of particles by the
Fokker-Planck equation\cite{risken}.
The basis as well as the limitation of this model is the assumption
that the process governing the non-Liouvillean effects is Markovian.
The conditions under which the Fokker-Planck approach can be applied to the
realm of ion optics has been discussed in earlier papers\cite{bohn,struck}.

We will try to extend this ansatz in the following by introducing
an entropy\cite{hobson} in a way that directly
relates it to the $\mu$-phase space density function.
This quantity will then serve as a means to identify beam dynamics
phenomena that are inherently irreversible and are hence associated
with an increase of entropy.
It will be shown in section~\ref{liouville} that the so-defined
entropy remains conserved as long as the $\mu$-phase
space Liouville theorem is fulfilled.
Entropy changes thus directly reflect the occurrence of non-Liouvillean
effects --- which in turn will be described by the Fokker-Planck equation.
This is the basis on which, in section~\ref{fokker}, the time derivative
of the entropy will be calculated assuming that the beam's
velocity distribution is Maxwellian.

For a special class of Markov processes --- namely for so-called
Ornstein-Uhlenbeck processes\cite{risken} ---
this expression may be considerably simplified.
As demonstrated in section~\ref{fluctdiss}, only {\em constant\/}
Fokker-Planck coefficients are then contained in the equation
for the change of entropy.
We may restrict ourselves to the Ornstein-Uhlenbeck model if
the non-Liouvillean part of the dynamics is small
compared to that conserving the $\mu$-phase space volume.
This is always true for charged particle beam optics.

If the Fokker-Planck coefficients appertaining to each degree of
freedom do not differ significantly from each other so that
the diffusion and friction processes
can be regarded as approximately isotropic, we may set up
the fluctuation-dissipation theorem in its simplest form.
This was first done by Einstein\cite{einstein},
in his work on the Brownian motion of particles.
It is obviously valid in situations not too far
from a fictitious thermodynamic equilibrium.
We will show that entropy growth is then directly related to heat
transfers between different degrees of freedom within the beam.

In section~\ref{rmsemi} we switch back to a more general treatment
and we first of all review the idea of a moment analysis of the
Vlasov-Fokker-Planck equation\cite{lapostolle,sacherer,struck}.
We are then prepared to set up the equation that relates entropy
growth to changes of the rms-emittance in conjunction with
changes of the ``excess field energy''.
In this context, the known equation relating changes of the
rms-emittance to changes of the ``excess field energy'' ---
first derived by Wangler\cite{wang} --- appears as a special case.

In the last section, the entropy equation is applied to
$1$- and $2$-dimensional beam models commonly used as simplifying
concepts in the theory of charged particle beams.
The well known phenomenon that simulations of periodic quadrupoles channels
always exhibit --- in contrast to solenoid channels --- a specific increase
of the rms-emittance is then easily explained and identified
as a computer noise artifact.
Furthermore, a scaling law for intra-beam scattering effects in real
($3$-dimensional) beams is obtained.
It states that the emittance growth rates are determined both by the
general beam parameters as well as by the particular optics of the beam
transport system.
\section{THE ENTROPY CONCEPT}\label{entro}
We start our analysis by defining the $6$-dimensional
normalized $\mu$-phase space density function
$$
f=f(x,y,z,p_x,p_y,p_z;t)\;.
$$
The quantity $f\,d^3q\,d^3p$ then represents the probability
of finding a particle inside a volume $d\tau=d^3q\,d^3p$ around
the phase space point $(x,y,z,p_x,p_y,p_z)$ at time $t$.
Following Shannon\cite{shannon}, the related information entropy
$S$ is given by
\begin{equation} \label{entro-def}
S(t)=-k_B\int f\,\ln f \,d\tau\;,
\end{equation}
with $f$ the $6$-dimensional $\mu$-phase space density function.
The change of the so-defined entropy, hence the time derivative
of Eq.~(\ref{entro-def}), then follows as:
\begin{equation} \label{entro-prod}
\frac{d S}{d t} = -k_B\int (1+\ln f) \,
\frac{\partial f}{\partial t} \,d\tau \;.
\end{equation}
This definition of an entropy does not contain any
resolution-dependent features.
If we could manage to measure directly the $\mu$-phase space
entropy, we would obtain the value of $S$ as defined in
Eq.~(\ref{entro-def}) only in the limit of perfect resolution.
In contrast, a real measuring device would provide us with reduced
amount of information on the $\mu$-phase space density function.
We could expect the measured coarse grained entropy $S_{\mathrm{cg}}$
to be larger than the true entropy $S$.
In any case, the respective value of $S_{\mathrm{cg}}$ depends
largely on the specific resolution of a measuring device.
It is thus not suited for a general purpose analysis.
We will therefore use the entropy definition (\ref{entro-def})
throughout in this article, keeping in mind that it only
constitutes an {\em idealized\/} entropy with regard to finite
resolution measuring devices.
\section{LIOUVILLEAN DYNAMICS}\label{liouville}
It can easily be shown that the total time derivative
of the $\mu$-phase space density function $f$ vanishes,
i.e.\ the $\mu$-phase space Liouville theorem applies
if the particles do not interact and if the time evolution
of their coordinates follows Hamilton's canonical equations.
For charged particle beams whose self-fields must be taken
into account, Liouville's theorem for the $\mu$-space $f$
remains fulfilled if the self-fields can be
treated analogously to the external focusing fields.
Explicitly, $d f / d t = 0$ leads to
\begin{equation} \label{liou-expl}
\frac{\partial f}{\partial\,t} = \sum_{i=1}^3 \left(
\frac{\partial f}{\partial p_i} \frac{\partial H}{\partial x_i} -
\frac{\partial f}{\partial x_i} \frac{\partial H}{\partial p_i}\right) \;.
\end{equation}
Inserting Eq.~(\ref{liou-expl}) into (\ref{entro-prod}), we
obtain after integration by parts
$$
\frac{d S}{d t} = 0 \;.
$$
Here we have made the reasonable physical assumption that the
phase space density $f$ as well as all its derivatives
vanish at the boundaries of the populated phase space.
Consequently, all integrated expressions
evaluate to zero at the integration boundaries.
Summarizing the above result, we may write
\begin{equation}\label{liou-entr}
\frac{d f}{d t} = 0 \quad \Longrightarrow \quad \frac{d S}{d t} = 0\;,
\end{equation}
i.e.\ the entropy change vanishes as long as Liouville's theorem
applies for $f$.
Liouville's theorem for the $\mu$-space $f$ does not apply
if --- for example --- particle-particle interactions
(``intra-beam scattering'') take place.
Thus if the actual ``granularity'' of the charge distribution
must be taken into account, we can no longer assume that
the ``single-particle'' distribution function $f$, i.e.\ the
lowest order of the ``BBGKY hierarchy''\cite{lawson},
contains all necessary data on the actual beam.

According to (\ref{liou-entr}), entropy changes are directly
related to violations of the $\mu$-phase space Liouville theorem.
An increase of entropy just implies an increase of the
$\mu$-phase space volume that the beam occupies and
hence an absolute degradation of the beam quality.
Processes that cause a phase space filamentation while {\em conserving\/}
the $\mu$-phase space Liouville theorem do {\em not\/}
change the entropy $S$, as defined in Eq.~(\ref{entro-def}).
Surely, such a phase space filamentation means a loss of
beam quality in a practical sense due to a lack of means
to reestablish the original phase space state.
Nevertheless, this process is not reflected by our definition
of the entropy since in the infinite resolution limit
a filamentation does not mean any loss of information.
\section{ENTROPY CHANGE ASSOCIATED WITH A MARKOV PROCESS}\label{fokker}
A precise analysis of effects that are due to the ``granularity''
of the charge distribution requires taking into account
the phase space coordinates of {\em individual\/} particles.
Obviously this kind of problem can never be tackled on the basis
of the deterministic approach embodied in the Vlasov equation.
On the contrary a {\em stochastic\/} contribution to the
net forces acting on a particle must be added.
In analogy to the Fokker-Planck description of the Brownian
motion of particles, we may model the action of random forces
within a beam by a process whose state at time $t+\Delta t$
depends only on its state at time $t$ and {\em not\/} on earlier times.
Here $\Delta t$ denotes a characteristic time interval that must
be small as compared to the time scale of macroscopic changes of the system.
A stochastic process possessing this property is usually referred to
as a Markov process.
It is easily shown\cite{jansen} that the equation of motion of such
processes is given by the Fokker-Planck equation.
The description of particle dynamics including stochastic forces can
thus be based on the combined Vlasov-Fokker-Planck equation\cite{chandra}
\begin{equation} \label{v-fp}
\frac{\partial\,f}{\partial\,t} + \frac{\vec{p}}{m} \cdot
\vec{\nabla}_x f + \left( \vec{F}^{\mathrm{ext}} +
q\vec{E}^{\mathrm{sc}} \right) \cdot \vec{\nabla}_p f =
{\left[ \frac{\partial\,f}{\partial\,t}\right]}_{\mathrm{FP}} \;,
\end{equation}
with
$$
{\left[ \frac{\partial f}{\partial t} \right]}_{\mathrm{FP}} \!\!\!\!\!\!=\!
- \!\sum_i \frac{\partial}{\partial p_i} \left[ F_i(\vec{p},t) \, f
\right] + m^2 \!\sum_{i,j} \frac{\partial^2}{\partial p_i \partial p_j}
\!\left[ D_{ij}(\vec{p},t) \, f \right].
$$
In this notation, $\vec{F}^{\mathrm{ext}}$ stands for the applied
external focusing forces, $q\vec{E}^{\mathrm{sc}}$ for the
macroscopic electric space charge forces, $F_i(\vec{p},t)$ for
the ``drift vector'' components of the Fokker-Planck equation, and
$D_{ij}(\vec{p},t)$ for its ``diffusion tensor'' elements.
As shown in the previous section, the Vlasov terms do not
contribute to any entropy production.
Therefore, Eq.~(\ref{entro-prod}) can be rewritten as
\begin{equation} \label{entro-prod2}
\frac{d S}{d t} = -k_B\int (1+\ln f) {\left[
\frac{\partial f}{\partial t} \right]}_{\mathrm{FP}} \,d\tau \;.
\end{equation}
Explicitly this means
\begin{eqnarray} \label{expli}
\frac{d S}{d t} = k_B\int (1+\ln f) &&\left\{
\sum_i \frac{\partial}{\partial p_i} \left[ F_i(\vec{p},t) \, f \right] -
\right.\nonumber\\
&&\left.\!\!\!\!\! m^2 \sum_{i,j} \frac{\partial^2}{\partial p_i \partial p_j}
\left[ D_{ij}(\vec{p},t) \, f \right] \right\} \,d\tau \;.
\end{eqnarray}
Integrating the terms of the first sum twice by parts, we obtain
$$
\int (1+\ln f) \, \frac{\partial}{\partial p_i} [ F_i f ] \,d\tau =
\int \frac{\partial F_i}{\partial p_i} f \,d\tau \;.
$$
Again we take advantage of the fact that for real beams
the phase space density $f$ as well as all
its derivatives vanish at the integration boundaries.

It has been shown by Reiser\cite{reiser} that the Maxwell-Boltz\-mann
distribution is the only one that provides a steady-state
solution of both, the time-independent Vlasov equation and
the time-independent Fokker-Planck equation.
We conclude that this distribution is best suited for
the description of a ``steady state beam'', i.e.\ a beam
which has adapted itself to the focusing structure.
Since the applied external forces vary along that structure, a
charged particle beam can never completely settle down to equilibrium.
Therefore, the instantaneous velocity distribution of a real beam must
be approximated by a non-isotropic Maxwell-Boltzmann distribution
that generalizes the steady state idealization
\begin{equation}\label{M-B-dist}
f\!\!=\!\!g(x,y,z;t) \exp \left( -\frac{p_x^2}{2m k_B T_x}
-\frac{p_y^2}{2m k_B T_y}-\frac{p_z^2}{2m k_B T_z}\right),
\end{equation}
with $g(x,y,z;t)$ as the self-consistent charge density and
the exponential function describing the distribution of the
incoherent part of the kinetic particle energy.
The coherent part of the kinetic energy of the beam particles
--- which originates in the ``breathing'' of the beam envelopes ---
can be eliminated since it does not cause any entropy changes.
We may therefore restrict ourselves in Eq.~(\ref{M-B-dist}) to a
principle axes formulation even for the case of a strong focusing and
dispersive system.

With the phase space density function~(\ref{M-B-dist}),
the terms of the second sum of Eq.~(\ref{expli}) evaluate to
$$
m^2 \int (1+\ln f) \, \frac{\partial^2}{\partial p_i \partial p_j}
[ D_{ij}f ] \,d\tau =-\frac{m}{k_{B}T_i}\delta_{ij}\int D_{ij} f\,d\tau\;.
$$
In summary, the change of entropy caused by a Markov process
can be expressed in terms of the Fokker-Planck coefficients as
\begin{equation} \label{entmark}
\frac{d S}{d t} = k_B \cdot \sum_{i=1}^3 \left( \< \frac{\partial F_i}
{\partial p_i} \> + \frac{m}{k_{B}T_i} \< D_{ii} \> \right)\;,
\end{equation}
wherein the angle brackets denote the respective averages
over the $\mu$-phase space density function $f$.
\section{ORNSTEIN-UHLENBECK PROCESSES}\label{fluctdiss}
The Fokker-Planck model --- as expressed mathematically in
Eq.~(\ref{v-fp}) --- is based on the assumption that the action
of the stochastic components of the interaction forces can be
described in terms of a diffusion process in velocity space
that is opposed by a dynamical friction force.
If these stochastic contributions to the dynamics of a system
are small, we may restrict ourselves to a subset of Markov processes,
referred to as Ornstein-Uhlenbeck processes\cite{ornuhl}.
The latter are defined by the property that its
Fokker-Planck equation contains a linear drift coefficient
together with a constant diffusion coefficient
\begin{equation} \label{o-u-coeff}
F_i = -\beta_{f;i}\cdot p_i \qquad,
\qquad \beta_{f;i}, D_{ii} = \mathrm{const.}
\end{equation}
This ansatz corresponds to Stokes's friction law in classical mechanics.
It applies to cases where the friction forces are small in comparison
to all other forces relevant for the dynamics of the system.
This is true in our context, since taking into account friction effects
among the beam particles always plays the role of a small correction.
Therefore, Eq.~(\ref{entmark}) simplifies to
\begin{equation} \label{entorn}
\frac{d S}{d t} = k_B \cdot \sum_i \left( -\beta_{f;i} +
  \frac{m}{k_B T_i} D_{ii} \right) \;.
\end{equation}
Eq.~(\ref{entorn}) forms the basis for establishing a relation
between entropy and rms-emittance, as will be shown in the next section.

At this point it is interesting to consider the special case
of isotropic Fokker-Planck coefficients.
This is surely correct for situations not too far from
a fictitious thermodynamic equilibrium where
the diffusion as well as the friction processes
can be treated as being approximately isotropic.
Eq.~(\ref{entorn}) then becomes
$$
\frac{d S}{d t} = k_B \cdot \sum_i \left( -\beta_f +
\frac{m}{k_B T_i} D \right) \;.
$$
The diffusion process arising from the fluctuations of the self-fields
and the friction effects associated with particle-particle interactions
are {\em not\/} independent of each other.
On the contrary, the diffusion coefficients $D_{ii}$ are related to the
friction terms $\beta_{f;i}$ via a fluctuation-dissipation theorem.
In the simplest case of an isotropic process, this
theorem is embodied in the Einstein relation\cite{einstein}
$$
D = \beta_f\, \frac{k_B T}{m} \;,
$$
wherein $T= \frac{1}{3}\sum_i T_i$ stands for the equilibrium temperature.
The entropy change due to a temperature balancing
process may then be written as
\begin{equation} \label{entiso}
\frac{d S}{d t} = k_B\beta_f\cdot\sum_i\left( \frac{T}{T_i} - 1\right)\;,
\end{equation}
or, explicitly
\begin{equation} \label{entisoexpl}
\frac{d S}{d t} = \frac{1}{3} k_B \beta_f \left[
\frac{{(T_x -T_y )}^2}{T_x T_y} + \frac{{(T_x -T_z )}^2}{T_x T_z} +
\frac{{(T_y -T_z )}^2}{T_y T_z} \right].
\end{equation}
Obviously, the entropy $S(t)$ remains unchanged in the case of
temperature equilibrium while increasing during temperature balancing:
$$
\frac{d S}{d t} \;\;
\begin{cases}
\;\;&=0\qquad\text{for temperature equilibrium}\\
\;\;&>0\quad\text{during temperature balancing}.
\end{cases}
$$
The total heat exchange $d Q/ d t$ vanishes,
as is easily seen from Eq.~(\ref{entiso})
\begin{equation}\label{heat}
\frac{d Q}{d t} \equiv \sum_i T_i \frac{d S_i}{d t} = k_B \beta_f \sum_i
\left( T- T_i \vphantom{T^T} \right) \equiv 0 \;.
\end{equation}
If we exclude effects such as radiation damping or dissipation of
electro-magnetic energy in the surrounding structure and assume that
no external heating or cooling devices are active,
this vanishing of the total heat exchange
is not surprising since a charged particle beam cannot
exchange heat with the focusing lattice.
Within the beam, heat exchange between the degrees of freedom may occur,
leading to an entropy growth as described by Eq.~(\ref{entisoexpl}).
We conclude that equipartitioning effects occurring within
initially thermally unbalanced charged particle beams are
always associated with an irreversible degradation
of the beam quality as a whole.
Furthermore, instantaneous temperature differences may exist
even if the beam is perfectly matched in all its moments on the
average over one focusing period.
In beam transport systems with quadrupole focusing,
apart from isolated locations, the instantaneous transverse
temperatures are always different.
Therefore, a certain growth rate --- depending on the size of
the temperature differences --- can never be avoided.

The time scale for this process is determined by the frequency $\beta_f$.
As the result of averaging procedures\cite{jansen,reiser}, this
follows from the global beam parameters as:
\begin{equation} \label{friction}
\beta_f = \frac{16 \sqrt{\pi}}{3} \; n \, c
{\left( \frac{q^2}{4\pi\epsilon_0 m c^2} \right)}^2 \cdot
{\left( \frac{m c^2}{2k_{B}T} \right)}^{3/2} \cdot \; \ln \Lambda \;.
\end{equation}
In this equation, $n$ stands for the real space average particle
density and $\ln \Lambda$ for the Coulomb logarithm.
\section{ENTROPY AND RMS-EMITTANCE}\label{rmsemi}
A second order moment analysis of the generalized Liouville equation
(\ref{v-fp}) yields the following set of coupled equations of
motion\cite{struck} for each phase space plane $i = 1,2,3$:
\begin{eqnarray} \label{deri}
\frac{d}{d t} \<x_i^2\> - \frac{2}{m} \<x_i p_i\> & = & 0 \\
\frac{d}{d t} \<x_i p_i\> - \frac{1}{m} \<p_i^2\> -
\<x_i F^{\mathrm{ext}}_i\> - q\, \<x_i E_i\> & = & \<x_i F_i \> \nonumber \\
\frac{d}{d t} \<p_i^2\> - 2\, \<p_i F^{\mathrm{ext}}_i \> -
2q\, \<p_i E_i\> & = & \nonumber\\
2\, \<p_i F_i \> & + & 2m^2\< D_{ii} \>  \nonumber \;,
\end{eqnarray}
with $F^{\mathrm{ext}}_i$ the components of the external focusing forces.
Again, the angle brackets enclose the respective averages over the
phase space density function: $\< a\>=\int a f d\tau$.
Using canonical variables, the rms-emittance in the beam system
is usually defined as
\begin{equation} \label{epsrms}
\varepsilon_{i,\mathrm{rms}}^2(t)\, =\,\<x_i^2\>\<p_i^2\> -\<x_i p_i\>^2\;.
\end{equation}
Since no other definitions of emittance are used throughout this
article, we will skip the index ``rms'' in the following.
On calculating the time derivative of $\varepsilon_i^2(t)$,
we readily obtain
\begin{eqnarray} \label{epsrms1}
\frac{d }{d t}\varepsilon_i^2(t) &=&
2\left[ \<x_i^2\> \<p_i F^{\mathrm{ext}}_i \> - \<x_i p_i\>
\<x_i F^{\mathrm{ext}}_i\> \right] + \nonumber\\ \vphantom{\frac{x}{t}}
&& 2 q\left[ \<x_i^2\>  \<p_i E_i\> -
\<x_i p_i\> \<x_i E_i\>\right] + \nonumber\\
&& 2\left[ \<x_i^2\> \<p_i F_i \> - \<x_i p_i\> \<x_i F_i\>\right] +\nonumber\\
&&2m^2 \<x_i^2\> \< D_{ii}\> \vphantom{\frac{x}{t}}.
\end{eqnarray}
The terms containing the external field components cancel if these fields
can be regarded as linear
\begin{equation}\label{linext}
F^{\mathrm{ext}}_i \propto x_i \Longleftrightarrow
\<x_i^2\>\<p_i F^{\mathrm{ext}}_i \>\equiv
\<x_i p_i\> \<x_i F^{\mathrm{ext}}_i\>\;.
\end{equation}
The so-called ``excess field energy'' --- namely the difference between
the field energy $W$ of an {\em arbitrary\/} charge distribution
and the field energy $W^{\mathrm{u}}$ of a {\em uniform\/} charge
distribution of the same rms-size --- is given by\cite{wang,stho}
\begin{equation} \label{dwdwu}
\frac{d}{d t} \left( W - W^{\mathrm{u}} \right) =
-\frac{N q}{m} \sum_i \left( \<p_i E_i\> -
\frac{\<x_i p_i\>}{\<x_i^2\>}\<x_i E_i \> \right) \;.
\end{equation}
We observe that the terms of the sum in Eq.~(\ref{dwdwu}) exactly
correspond to the moments involving the electric self-fields
$E_i$ in (\ref{epsrms1}).

Assuming again that the non-Liouvillean process can be approximated
by an Ornstein-Uhlenbeck process, we may simplify the
Fokker-Planck coefficients according to Eq.~(\ref{o-u-coeff}).
Together with Eqs.~(\ref{linext}) and (\ref{dwdwu}), the equation
of motion for the rms-emittance (\ref{epsrms1}) can be rewritten as
\begin{equation} \label{epsrms2}
\frac{1}{\<x_i^2\>} \frac{d}{d t}\varepsilon_i^2(t) \!=\!
-2\beta_{f;i} \frac{\varepsilon_i^2(t)}{\<x_i^2\>} \!+\!
2 m^2 D_{ii} \!-\!\frac{2m}{N}\frac{d}{d t} \left( W_i -
W_i^{\mathrm{u}} \right),
\end{equation}
with $W_i - W_i^{\mathrm{u}}$ denoting formally the
$i$-th component of the sum~(\ref{dwdwu}).

The global ``temperature'' $T_i$ of the $i$-th degree of freedom of a
charged particle beam can be expressed in terms
of second order beam moments
\begin{equation}\label{temp}
k_B T_i = \frac{1}{m} \frac{\varepsilon_i^2(t)}{\<x_i^2\>} \;,
\end{equation}
provided that the projections of the phase space density function $f$
onto $2$-dimensional subspaces $(x_i,p_i)$ are homogeneously populated.
To a good approximation, this expression can be applied to arbitrary
phase space density functions $f$ since the respective error always
vanishes at the extremities of the beam envelopes\cite{struck}.

With the help of this approximation, the temperature $T_i$
contained in the equation for change of entropy (\ref{entorn})
can be replaced by the corresponding beam moments
\begin{equation}\label{entmoment}
\frac{d S}{d t} = \sum_i \frac{d S_i}{d t} \quad ,\quad
\frac{d S_i}{d t} = k_B \left( -\beta_{f;i} +
m^2 \frac{\<x_i^2\>}{\varepsilon_i^2(t)} D_{ii} \right) \;.
\end{equation}
Inserting Eq.~(\ref{entmoment}) into (\ref{epsrms2}), we obtain an
equation relating emittance, entropy and excess field energy
\begin{equation}\label{emi-ent-1}
\frac{1}{\<x_i^2\>} \frac{d}{d t}\varepsilon_i^2(t) =
\frac{2}{k_B} \,\frac{\varepsilon_i^2(t)}{\<x_i^2\>} \,\frac{d S_i}{d t} -
\frac{2m}{N}\frac{d}{d t} \left( W_i - W_i^{\mathrm{u}} \right) \;.
\end{equation}
As the last step, the summation over $i$ must be performed
\begin{eqnarray}\label{ent-fe-1}
\sum_i \frac{1}{\<x_i^2\>} \frac{d}{d t}\varepsilon_i^2(t) &+&
\frac{2m}{N}\frac{d}{d t} \left( W - W^{\mathrm{u}} \right) =\nonumber\\
&=&2m \,\sum_i T_i \frac{d S_i}{d t} \mathrm{\;\;\;in\;general}\\
&=&0 \mathrm{\;\;\;for\; isotropic\; FP\; coefficients\;.} \nonumber
\end{eqnarray}
As stated before in Eq.~(\ref{heat}), the right hand side of
(\ref{ent-fe-1}) sums up to zero under the precondition of
isotropic Fokker-Planck coefficients.
Eq.~(\ref{ent-fe-1}) then constitutes the known relationship
between the changes of the rms-emittances and the change of
the excess field energy, first derived by
Wangler\cite{wang,host,stho} in a pure Vlasov approach.
As we learn now, this equation even holds
if Liouville's theorem in the $\mu$-phase space does not apply
as long as the non-Liouvillean effects can approximately
be described by isotropic Fokker-Planck coefficients.

Multiplying  Eq.~(\ref{emi-ent-1}) with
$\<x_i^2\>/ 2\varepsilon_i^2(t)$ leads to the equivalent form
\begin{equation}\label{emi-ent-2}
\frac{1}{k_B} \frac{d S_i}{d t} =
\frac{d}{d t} \ln \varepsilon_i(t) +
\frac{m}{N} \, \frac{\<x_i^2\>}{\varepsilon_i^2(t)}\,
\frac{d}{d t} \left( W_i - W_i^{\mathrm{u}} \right) \;.
\end{equation}
Summing now Eq.~(\ref{emi-ent-2}) over $i$,
the time derivative of the entropy function $S(t)$ becomes
\begin{eqnarray}\label{ent-fe-2}
\frac{1}{k_B} \,\frac{d S}{d t} & = &
\frac{d}{d t} \ln \varepsilon_x(t)\varepsilon_y(t)\varepsilon_z(t) \!+\!
\frac{m}{N} \!\sum_i \!\frac{\<x_i^2\>}{\varepsilon_i^2(t)}
\frac{d (W_i-W_i^{\mathrm{u}})}{d t}\nonumber\\
& = & \frac{d}{d t} \ln \varepsilon_x(t)\varepsilon_y(t)\varepsilon_z(t) -\nonumber\\
&&\qquad q\sum_i \frac{\<x_i^2\> \<p_i E_i\> - \<x_i p_i\> \<x_i E_i\>}
{\varepsilon_i^2(t)}\;.
\end{eqnarray}
This equation constitutes a general relation between entropy
change, the change of the rms-emittances, and the temperature
weighted change of the excess field energy for the realm of ion optics.
It thus confirms the heuristic approach presented earlier by
Lawson~et~al.\cite{lalaglu}, who showed the close relation
between the entropy and beam emittance.

With the heat differential $d Q_i$ defined as
$$
dQ_i = k_B T_i \,d \ln \varepsilon_i +
\frac{1}{N} \,d \left( W_i-W_i^{\mathrm{u}}\right)\;,
$$
Eq.~(\ref{ent-fe-2}) reads
$$
dS = \sum_i \frac{d Q_i}{T_i} \;,
$$
using the temperature definition of Eq.~(\ref{temp}).
For the special case of isotopic Fokker-Planck
coefficients, the emittance to excess field energy
relation (\ref{ent-fe-1}) then takes on the simple form
$$
\sum_i d Q_i = 0 \;.
$$
We note that Eq.~(\ref{emi-ent-1}) as well as Eq.~(\ref{emi-ent-2})
do not contain any Fokker-Planck coefficients --- although they are
derived on the basis of the Fokker-Planck approach~(\ref{v-fp}).
Recalling Eq.~(\ref{epsrms2}), we see that the Fokker-Planck
related moments exactly agree with those appearing in Eq.~(\ref{entmark}) ---
provided that we restrict ourselves to Ornstein-Uhlenbeck processes, and
the global temperature definition (\ref{temp}).
Under these preconditions, the insertion of Eq.~(\ref{entmoment}) into
Eq.~(\ref{epsrms2}) leads to a complete replacement of all terms containing
Fokker-Planck coefficients by the function for the change of entropy.
The Fokker-Planck approach is thus included in Eqs.~(\ref{emi-ent-1})
and (\ref{emi-ent-2}) just by allowing for
changes of the entropy (\ref{entro-def}),
and {\em not\/} by eliminating entropy changes $d S_i$ a priori,
as it is done in a Vlasov approach.

In the course of this derivation, the temperatures $T_i$,
as defined in Eq.~(\ref{temp}), are understood as global
temperatures pertaining to the $i$-th degree of freedom.
Implicitly, we thus assumed that
no heat is transferred within each degree of freedom.
In other words we only treat cases where the beam has already
adapted itself to the focusing lattice, i.e.\
cases where no transient effects are observed.
This condition is not necessarily fulfilled.
It has been shown numerically by various authors
(cf.~\cite{lapostolle,strklarei,anderson}, for example) that
a redistribution of the populated phase space --- occurring
if a beam is launched with a non-self-consistent phase space filling ---
also constitutes an irreversible process.
These effects are not covered by our approach since
for non-self-consistent phase space densities,
a transfer of heat also takes place within each degree of freedom.
A global temperature description is not sufficient
under these circumstances.
On the contrary, a local, i.e.\  spatially dependent
``temperature'' definition must be used instead.
\section{DISCUSSION}\label{discus}
\subsection{$1$-D beam model}
For the sake of mathematical simplicity, the
sheet beam model is sometimes applied, since
it allows analytical solutions for cases where
more realistic models depend on numerical methods.
Of course, the one-dimensional beam model is oversimplified
in the sense of not allowing heat transfer to other
degrees of freedom.
With regard to the derivations of the last section, this means
that no summation over $i$ must be performed.
In other words, because of this model, Eqs.~(\ref{ent-fe-1})
and (\ref{ent-fe-2}) are equivalent
$$
\frac{1}{k_B} \,\frac{d S}{d t} =
\frac{\<x^2\>}{2\varepsilon_x^2(t)}\left[
\frac{1}{\<x^2\>} \frac{d}{d t}\varepsilon_x^2(t) +
\frac{2m}{N}\frac{d}{d t}\left( W -W^{\mathrm{u}}
\right)\right] \!\equiv 0\,.
$$
Consequently, all solutions of the Vlasov-Fokker-Planck e\-qua\-tion
(\ref{v-fp}) are reversible if the initial phase space density
function is intrinsically matched.
This behavior of sheet beams has been described and numerically
simulated earlier by Anderson\cite{anderson}, who showed the
existence of strictly reversible changes of the rms-emittance.
\subsection{$2$-D beam model}
The $2$-dimensional $x,y$-beam model is widely used in analytical as
well as in numerical approaches to the study of the transformation
of unbunched (``coasting'') beams.
With the equilibrium temperature $T=\frac{1}{2}(T_x+T_y)$ for the
$2$-D beam model, Eq.~(\ref{entiso}) for the entropy change near
thermodynamic equilibrium can be rewritten as
\begin{equation}\label{ent-2D}
\frac{d S}{d t} = \frac{1}{2} k_B \beta_f
\frac{{(T_x -T_y )}^2}{T_x T_y}\;.
\end{equation}
Beam transport without an increase of entropy (i.e.\
reversible beam transformations) are thus possible if either
\begin{itemize}
\item[(1)] $\beta_f \equiv 0$, which means that no non-Liouvillean
effects are present, or if
\item[(2)] $T_x \equiv T_y$, i.e.\ the beam stays round throughout
its propagation.
\end{itemize}
The first case just describes the pure Vlasov approach, which is
--- by our definition of the entropy in Eq.~(\ref{entro-def}) ---
always associated with a vanishing entropy growth, as already stated
in section~\ref{liouville}.

The second case states that no degradation of the beam quality
occurs, as long as no heat is transferred between the transverse
degrees of freedom.
This condition is met in the \mbox{2-D} beam model if we transform
a matched beam through a continuous or interrupted solenoid channel.

We note that with regard to intra-beam scattering effects, the heat
exchange with the longitudinal degree of freedom cannot be neglected.
In other words, the \mbox{2-D} beam model is not adequate for the
estimation of emittance growth rates due to intra-beam scattering.
This topic will be discussed in the next subsection.

\begin{figure}[ht]
\begin{center}
\epsfig{file=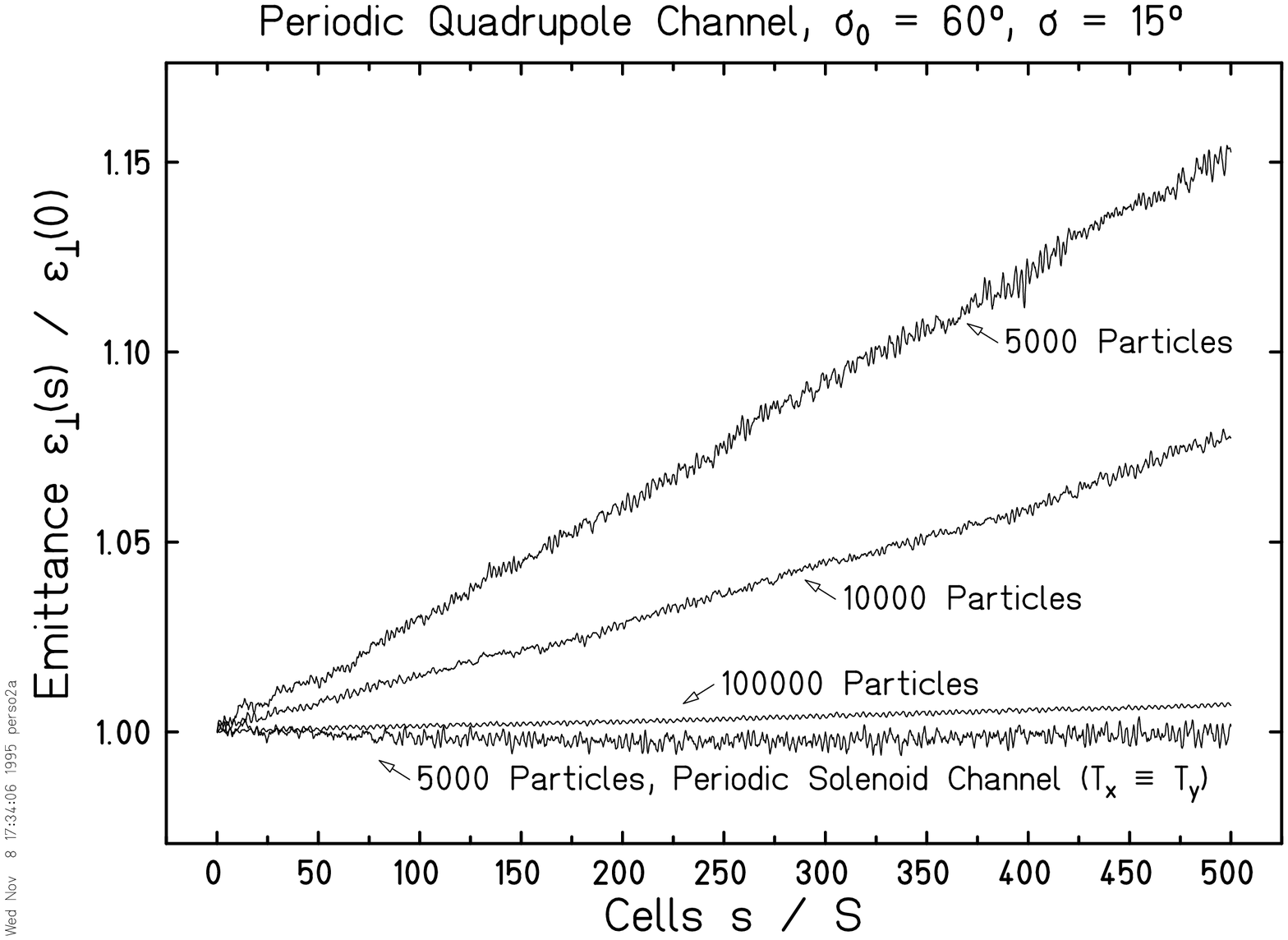,width=86mm,height=55mm}
\caption{Emittance growth factors versus number of cells obtained
by 2-D particle-in-cell simulations of beam transport channels
at $\sigma_0=60^\circ$, $\sigma=15^\circ$.
The upper three curves display the results of quadrupole
channel simulations with different numbers of simulation particles.
For comparison, the lowest curve shows the emittance growth
factors of a periodic solenoid channel simulation.}
\label{simu}
\end{center}
\end{figure}
\vspace*{-6mm}
Nevertheless, Eq.~(\ref{ent-2D}) can help us to interpret results of
computer simulations that are based on the 2-D beam model.
The upper three curves in Fig.~\ref{simu} show the evolution of the
rms-emittance growth factors along a quadrupole channel as they are
obtained for different numbers of macro-particles used in the simulation,
while keeping all other simulation parameters unchanged.
In all cases, the beam is launched with a self-consistent water-bag
distribution\cite{lapo,gluck,stho} for the initial phase space population.
The external focusing has been approximated by strictly linear
forces and the hard edge lens model.
The space charge fields have been determined using a fast
$x,y$-Poisson solver with $128$ by $128$ mesh nodes.
Under these circumstances only the space charge fields
can contribute to a growth of the rms-emittances.
Since the growth rates obtained as well as the amplitude of the
emittance fluctuations are approximately
inversely proportional to the number of macro-particles
used in the simulation, it is obviously the inaccuracies in calculating
these fields that are responsible for the growth of the rms-emittance.
With regard to Eq.~(\ref{ent-2D}), we conclude that these inaccuracies
induce a positive ``simulation friction coefficient''
$\beta_f^{\,\mathrm{sim}}$, which is to first order inversely
proportional to the number of simulation particles.

Using Eq.~(\ref{ent-2D}) to explain simulation results
means --- after all --- to use the Fokker-Planck equation (\ref{v-fp})
as the basis for the description of purely numerical noise phenomena.
The validity of this approach becomes obvious if we recall that
the gradual loss of information due to simplifications and roundoff
errors itself constitutes a Markov process which in turn can be modeled
by the Fokker-Planck equation.
This statement is confirmed by the simulation results
displayed in the lower curve of Fig.~\ref{simu}.
It shows the evolution of the rms-emittance growth factors during the
propagation of a matched beam through a periodic solenoid channel.
Since the beam stays round along the entire channel,
no transverse temperature gradient exists, hence no
entropy change is expected according to Eq.~(\ref{ent-2D}).
\begin{figure}[t]
\begin{center}
\epsfig{file=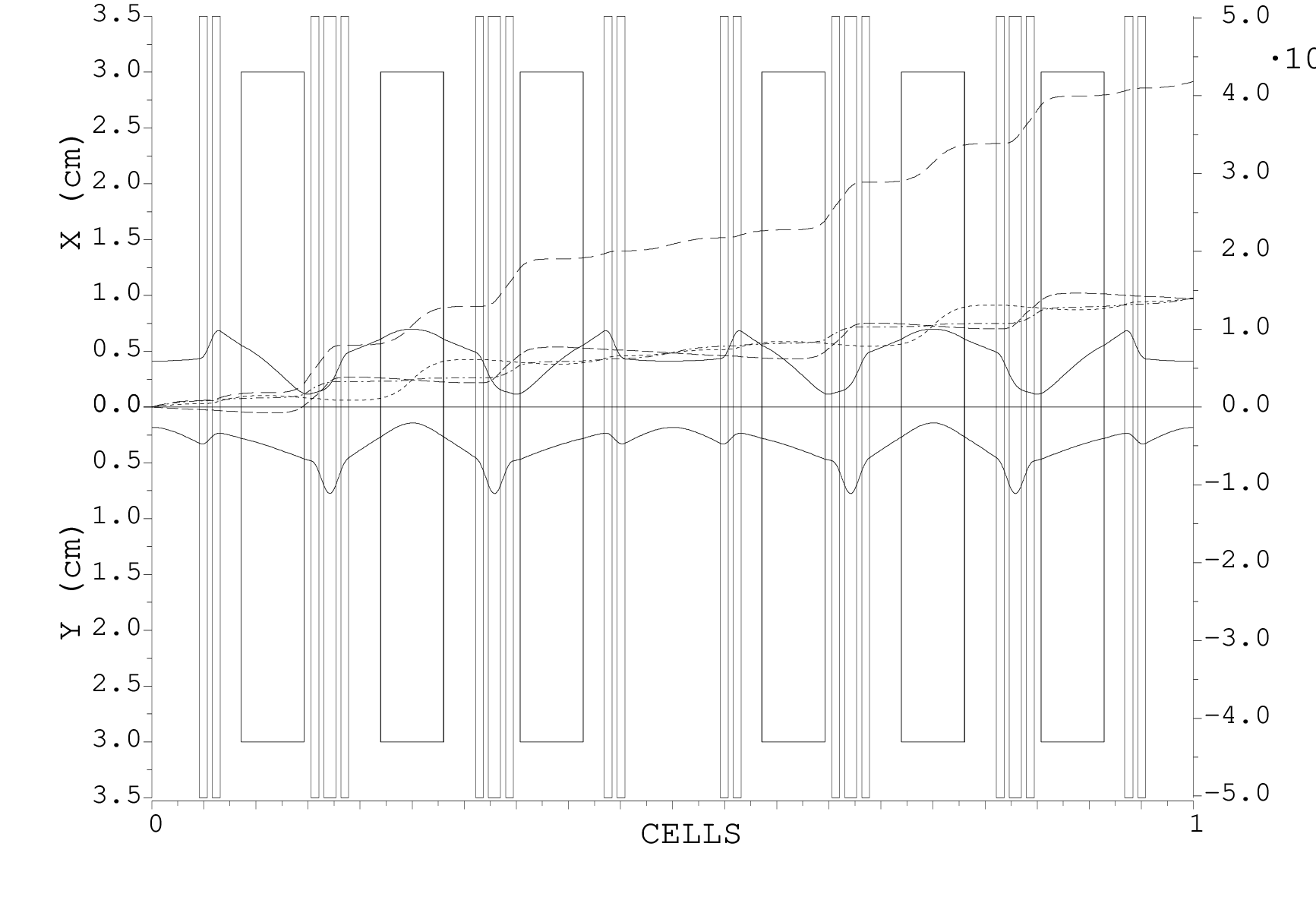,width=80mm,height=60mm}
\caption{Envelopes (solid lines), rms-emittance growth functions
($\varepsilon_x/\varepsilon_{x;0} - 1$: dotted line,
$\varepsilon_y/\varepsilon_{y;0} - 1$: dashed line,
$\varepsilon_z/\varepsilon_{z;0} - 1$: dashed-dotted line),
and entropy function $(S-S_0)/k_B$ (upper dashed line) of a
thermally matched beam passing through one turn of the GSI
Experimental Storage Ring (ESR) at $Q_h=2.31$ and $Q_v=2.25$.
The scale on the right hand side applies to the dimensionless
emittance and entropy growth functions.}
\label{entesr}
\end{center}
\end{figure}
We observe that the emittance fluctuations are similar in amplitude
to those in the quadrupole channel simulation performed with the
same number of macro-particles.
This means that in both cases the fluctuating part of the
self-fields impose a similar ``simulation friction coefficient''
$\beta_f^{\,\mathrm{sim}}$.
Yet, due to the lack of temperature differences, these fluctuations
do {\em not\/} produce an overall increase of the rms-emittance.
\subsection{3-D beam}
We first consider the hypothetical case of a beam that is thermally
balanced in all three dimensions, i.e.\ $T=T_x=T_y=T_z$.
As is easily seen, Eqs.~(\ref{ent-fe-1}) and (\ref{ent-fe-2}) are
again equivalent in this case, which in turn means that
the entropy remains constant.
If we imagine this beam is ``breathing'' isotropically in all three
directions, a completely reversible exchange between the rms-emittances
and the excess field energy (\ref{dwdwu}) would take place.
In this sense we may state that
\begin{itemize}
\item a change of the rms-emittance due to a change of the excess field
energy is a {\em reversible\/} process, whereas
\item a change of the rms-emittance due to a flow of heat is always
an {\em irreversible\/} process.
\end{itemize}
In real beam guiding systems, we must always cope with a specific
amount of temperature anisotropy and hence always deal
with a positive growth rate of the entropy.
The elementary mechanism responsible for the transfer of incoherent
kinetic energy from one degree of freedom to another
is constituted by the effect of Coulomb scattering of
individual beam particles.
This effect forms the basis for deriving Eq.~(\ref{friction}).
We may therefore use it in order to
estimate the time scale for heat flow effects, or --- equivalently ---
the scattering induced irreversible emittance growth rates.
In order to gain a better physical insight, the frequency $\beta_f$
can be expressed alternatively as
$$
\beta_f = \sqrt{\frac{2}{\pi}} \cdot {(\Delta
t_{\mathrm{scattering}})}^{-1}\cdot N\Gamma^2 \cdot \ln\Lambda\;\; ,
$$
with $ \Delta t_{\mathrm{scattering}}$ denoting the average time between
two successive scattering events of a beam particle,
$N$ the number of particles, and $\Gamma$ the
dimensionless coupling constant of the beam plasma.

If we neglect the (reversible) changes of the rms-emittance due
to changes of excess field energy, on the basis of Eqs.~(\ref{deri})
we may easily establish a closed coupled set of generalized envelope
and temperature change equations\cite{erlangen,struck}
which can be directly integrated.
The results of an integration of this set of equations based on the
geometry of the GSI experimental storage ring (ESR) are
plotted in Fig.~\ref{entesr}.
It includes the dispersion function
(``slip factor'')\cite{epac} for the particular tuning of the ring.
The ratios of the initial emittances have been optimized
to yield the same growth rates in all three dimensions.
The upper dashed curve thus displays
the minimum entropy growth $(S-S_0)/k_B$ during one turn.

Integrating Eq.~(\ref{entisoexpl}) we find
$$
\frac{1}{k_B}\big[S(t) - S_0\big] = \frac{1}{3}
\beta_f \, t\cdot \big[ I_{x y}(t) + I_{x z}(t) + I_{y z}(t) \big]\;,
$$
with
$$
I_{x y}(t) = \frac{1}{t}\;\int_{t_0}^t\;
\frac{{[T_x(t')-T_y(t')]}^2}{T_x T_y} \,d t'\;,
$$
and with $I_{x z}(t)$ and $I_{y z}(t)$ to be defined likewise.
If the global beam data --- thus $\beta_f$ --- is given, we can only
reduce the entropy production by minimizing the dimensionless sum
$$
I(t_1) = I_{x y}(t_1) + I_{x z}(t_1) + I_{y z}(t_1)\;,
$$
where $t_1$ denotes the time the beam centroid needs
to propagate over one focusing period.
For a given structure, this implies minimizing the average
temperature gradients and hence perfectly matching the beam
to the guiding structure in all three dimensions.
In our example (Fig.~\ref{entesr}) --- simulating the transformation
of a coasting beam through one turn of the ESR --- this has been
performed by rms-matching the transverse beam parameters
while at the same time adjusting the initial emittance
and momentum spread ratios.

If we are in the design phase of an ion optical system,
we may include minimizing of $I(t_1)$ as part of the
optimization of the structure under consideration.
This is just an alternative formulation of Reiser's
suggestion of a thermodynamic accelerator design\cite{reibro}.
\section{CONCLUSIONS}
We have used a resolution-independent entropy that follows directly
from the 6-dimensional $\mu$-phase space density function $f$.
This entropy possesses the property of remaining constant as long as
Liouville's theorem applies for $f$.
Since in reality as well as in computer simulations of charged
particle beams this is not true in a strict sense, we always
have to cope with a certain rate of dilution of the $\mu$-phase
space density and hence always accept some increase of entropy.
For the estimation of this growth, a new equation has been derived on
the basis of the Fokker-Planck approach.
It relates the change of entropy to the joint changes of the
rms-emittance and the temperature weighted excess field energy.
From this equation, we can conclude that the exchange of rms-emittance
and excess field energy may be performed without a change of entropy
and hence in a reversible manner.
In contrast to this, it has been shown that all heat transfers within
the beam --- feeding thermal energy from one degree of freedom to
another --- are always associated with an increase of beam entropy and
thus always lead to an irreversible degradation of beam quality.
\acknowledgments%
The author wishes to thank his GSI colleagues I.~Hofmann and C.~Riedel
for the valuable discussions during the course of this work.
He is also indebted to D.~Barber (DESY) for his critical
and thoughtful remarks.

\end{document}